\documentclass[fleqn,10pt]{wlscirep}
\usepackage[utf8]{inputenc}
\usepackage[T1]{fontenc}
\usepackage{bm}				 
\usepackage{hyperref}
\usepackage{etoolbox}
\usepackage{color} 
\usepackage{amsmath}

\newcommand{\mypsi}[1]{\psi_{#1}}
\newcommand{\mmypsi}[2]{\psi_{#1}^{#2}}

\title{Weakly-supervised learning on Schr\"{o}dinger equation}

\author[1,2,*]{Kenta Shiina}
\author[2,3,4,5,6]{Hwee Kuan Lee}
\author[1]{Yutaka Okabe}
\author[1]{Hiroyuki Mori}
\affil[1]{Department of Physics, Tokyo Metropolitan University, Hachioji, Tokyo, 192-0397, Japan}
\affil[2]{Bioinformatics Institute, Agency for Science, Technology and Research (A*STAR), 30 Biopolis Street, \#07-01 Matrix, 138671, Singapore, Singapore}
\affil[3]{School of Computing, National University of Singapore, 13 Computing Drive, 117417, Singapore, Singapore}
\affil[4]{Singapore Eye Research Institute (SERI), 11 Third Hospital Ave, 168751, Singapore, Singapore}
\affil[5]{Image and Pervasive Access Laboratory (IPAL), 1 Fusionopolis Way, \#21-01 Connexis (South Tower), 138632, Singapore, Singapore}
\affil[6]{Rehabilitation Research Institute of Singapore, 11 Mandalay Road \#14-03, Clinical Sciences Building, 308232, Singapore, Singapore}
\affil[*]{e-mainl: 16879316kenta@gmail.com}

\begin{abstract}
We propose a machine learning method to solve Schr\"{o}dinger equations for a Hamiltonian that consists of an unperturbed Hamiltonian and a perturbation.
We focus on the cases where the unperturbed Hamiltonian can be solved analytically or solved numerically with some fast way.
Given a potential function as input, our deep learning model predicts wave functions and energies using a weakly-supervised method. 
Information of first-order perturbation calculation for randomly chosen perturbations is used to train the model. 
In other words, no label (or exact solution) is necessary for the training, which is why the method is called weakly-supervised, not supervised. 
The trained model can be applied to calculation of wave functions and energies of Hamiltonian containing arbitrary perturbation. 
As an example, we calculated wave functions and energies of a harmonic oscillator with a perturbation and results were in good agreement with those obtained from exact diagonalization.
\end{abstract}

\begin{document}

\flushbottom
\maketitle
%
%
\thispagestyle{empty}

\section*{Introduction}
\label{sec:Intro}
Schr\"{o}dinger equation is a fundamental equation of quantum mechanics 
and its eigen functions (wave functions) contain information to analyze a target system.
However, Schr\"{o}dinger equation cannot be solved exactly in most cases and needs to be solved approximately or numerically.

There are many numerical techniques to solve Schr\"{o}dinger equations. 
Each of them have advantages and disadvantages and one usually verifies the accuracy of obtained results using different numerical methods 
which mutually compensate the disadvantages.
For example, exact diagonalization of a Hamiltonian is one of the most frequently used techniques to derive eigen functions of the Hamiltonian.
In addition, spectral method \cite{Livermore1982a}, density matrix renormalization \cite{Schollwock2005}, tensor network \cite{Orus2014}, 
and quantum Monte Carlo method \cite{Ahufinger2005} have been developed to investigate many-body systems.
Although these methods were successful to a certain degree in approximating solutions,
there are still difficulties when we deal with a system having a large numbers of degree-of-freedom, complex interactions, etc. 
Therefore, a new approach which can overcome these difficulties is desirable.

Deep Learning (DL) methods have been used as a mean to analyze data from various perspectives. 
With a large amount of data fed to DL, it extracts essential characteristics of the data and learns how to express them efficiently.
This feature of DL helps us predict and understand a behavior of a complicated system. 
In condensed matter physics, DL methods have been applied in various ways: 
classification of phase transitions \cite{PhysRevB.94.2016, Carrasquilla2017, PhysRevE.96.2017, PhysRevX.7.2017, jps2018, PhysRevB.97.2018, Shiina2020},
variational methods \cite{Carleo2017,Saito2018,HAN2019108929,PhysRevLett.124.020503,Manzhos_2020},
and efficiency improvement of conventional numerical methods \cite{PhysRevB.95.035105,Liu2017,Nomura2017,PhysRevLett.121.260601}.

DL methods have also been used to solve Schr\"{o}dinger equations 
as differential equations \cite{Mills2017, Liang2018, Raissi2019a, PhysRevA.98.010701},
following the development of DL solvers for differential equations \cite{Lagaris1998a, Sirignano2018, Raissi2018, Chen2018a}.
The objective of these studies is to design a DL solver for Schr\"{o}dinger equations which is applicable to various situations.
It is achieved by ``generalization,'' one of the important features of DL.
In order to obtain a generalized DL solver, we need to train DL by giving a large amount of data.

There are some machine learning paradigms such as supervised, unsupervised, weakly-supervised, and semi-supervised learnings
\cite{Bishop, Zhu, 10.1093/nsr/nwx106}. 
Due to the limitation of space, we briefly discuss the supervised, unsupervised and weakly-supervised paradigms in this paragraph.
In general, the supervised learning shows better generalization 
because mapping between data and associated labels (ground truth in the DL terminology) is fully provided.
However, preparing labels of data for training can sometimes be costly or even impossible.
On the other hand, the unsupervised learning does not require labels but the training can be difficult. 
Therefore, the unsupervised learning has a lower ability of generalization.
Finally, weakly-supervised learning is a paradigm that resides in between supervised and unsupervised learnings.
There are typically three types of weakly supervised learning;
incomplete supervision, where some instances in the training data lack labels;
inaccurate supervision, where the given labels are not always correct;
and inexact supervision, where the labels are not as exact as desired \cite{10.1093/nsr/nwx106}.
Our method is actually corresponding to a case of the inexact supervision.

In this paper, we propose a DL method to solve time-independent Schr\"{o}dinger equations (TISEs) with various types of potentials.
Several ingredients motivated us to use the DL approach. 
Firstly, quite realistic data related to the Schr\"{o}dinger equation can be generated efficiently. 
We can cheaply generate random potentials for the Hamiltonian of Schr\"{o}dinger equation, and DL can be used to train on these potentials. 
With a large amount of data, DL can converge to the exact solution effectively and predict the solutions for unseen potentials. 
The method of generating realistic random potentials will be described later in this paper. 
Secondly, DL computations for the solutions of Schr\"{o}dinger equation during inference are much more efficient 
than other methods such as exact diagonalization methods (see Supplementary Information).
Lastly, we propose a weakly-supervised method, in which we do not need expensive computations to provide the exact solutions (or ground truth) to train our DL method.
Instead, we make use of information about perturbations.
Therefore, the method can be applied to a case where we cannot obtain the ground truth or a case where it takes a long time to numerically obtain the ground truth.
Another advantage of our weakly-supervised method is that it is applicable to any quantum state if its perturbation calculation can be performed.

\section*{Method} 
\label{sec:method}
In this section, we first briefly introduce the perturbation theory,
a fundamental technique in quantum mechanics for finding an approximate solution of Schr\"{o}dinger equation.
We also explain our method of solving Schr\"{o}dinger equations by weakly-supervised machine learning based on the perturbation theory.
Through this paper, we consider one dimensional cases with coordinate variable $x\in\mathbb{R}$
although our formulation can be generalized to the case of arbitrary dimension.
We consider the case of arbitrary dimension for future work.

  \subsection*{Time-independent perturbation theory}
  Let $\mypsi{n}(x)$ be the n-th eigen function for Hamiltonian $H$ and $E_n$ be the corresponding eigen energy.
  Then, the one dimensional TISE for $H$ is given by
  \begin{equation}
   H\mypsi{n}(x) = E_n\mypsi{n}(x).
  \end{equation}
  Hereafter, when there is no ambiguity, we omit the variable $x$ in our equations to simplify notations.
  The basic idea of the perturbation theory is to calculate a solution by writing $H$ as a sum of unperturbed and perturbed terms. 
  The unperturbed term is $H_0$, whose eigen function $\mmypsi{n}{(0)}$ and eigen energy $E_n^{0}$ are assumed to be known,
  and the perturbation term is $\lambda\hat{V}$, where $\lambda \in \mathbb{R}$.
  If $\lambda \ll 1$, we can write down an eigen function $\mypsi{n}$ of $H$ 
  as the sum of an unperturbed term and correction terms in a power series of $\lambda$:
  \begin{equation}
   \label{eq:expandP}
   \mypsi{n} = \mmypsi{n}{(0)} + \lambda \mmypsi{n}{(1)} + \lambda^2 \mmypsi{n}{(2)} + \cdots.
  \end{equation}
  Here, the subscripts indicate the quantum number and the superscripts the perturbation order.
  The perturbation corrections can be calculated from the unperturbed  wave functions and energies. 

  This theory has been extremely successful in finding approximate solutions or analyzing effects of the perturbation
  although there are two major limitations on the theory. 
  One is that the perturbation theory can be adopted only when the introduction of the perturbation is adiabatic causing no discontinuity such as phase transition.
  The other is that, when $\lambda$ becomes larger, higher order terms need to be calculated to maintain the accuracy high enough.
  In other words, for large $\lambda$, the residual wave function defined as
  \begin{equation}
   \label{eq:res}
    \mmypsi{n}{(res)} \equiv \mypsi{n} - \mmypsi{n}{(0)}
  \end{equation}
  can be large and using only lower order terms results in poor approximation.

  In this paper, we focus on systems with perturbations which could induce only adiabatic changes.
  Our method can be well generalized and predict the residual wave functions and energies of an arbitrary quantum state $n$ for various perturbation potentials,
  only by using information of the first-order perturbation term with the help of DL. 

  \subsection*{Weakly-supervised}
  Let us consider a set of Hamiltonians, either of which consists of the same unperturbed term and a different perturbation term, ${H}_d = {H}_0 + {V}_d$.
  Here, the subscript $d$ distinguishes the perturbations, and ${V}_d={V}_d(x)$.
  We employ convolutional neural networks (CNNs) whose inputs are ${V}_d$
  and desired outputs are the residual wave function $\mmypsi{n,d}{(res)}$ and residual energy defined as $E_{n,d}^{(res)} \equiv E_{n,d} - E_{n}^{(0)}$.
  Here, $\psi_{n,d}$ and $E_{n,d}$ are the $n$-th eigen function and eigen energy of the Hamiltonian ${H}_d$.
  Our goal is making the CNNs learn the mapping from an arbitrary perturbation potential to a corresponding residual wave function 
  and energy without using exact solutions (i.e. ground truth).
  To achieve the goal, we prepare a variety of potentials and first-order corrections to respective energies as training data set. 
  By feeding them to the CNNs, the CNNs can learn how to solve TISE with the support of the information of the first-order energies.
  The trained CNNs can be generalized and give approximate solutions for different perturbation potentials which are not given in the training process.
  We call this method weakly-supervised learning because it does not require the exact solution provided by an expensive computation (e.g. diagonalization),
  but uses the first-order perturbation corrections to train the CNNs.
  Details of the method will be given in the following sections.

   \subsubsection*{Architecture}
\begin{figure}
    \centering
    \includegraphics[scale=0.5]{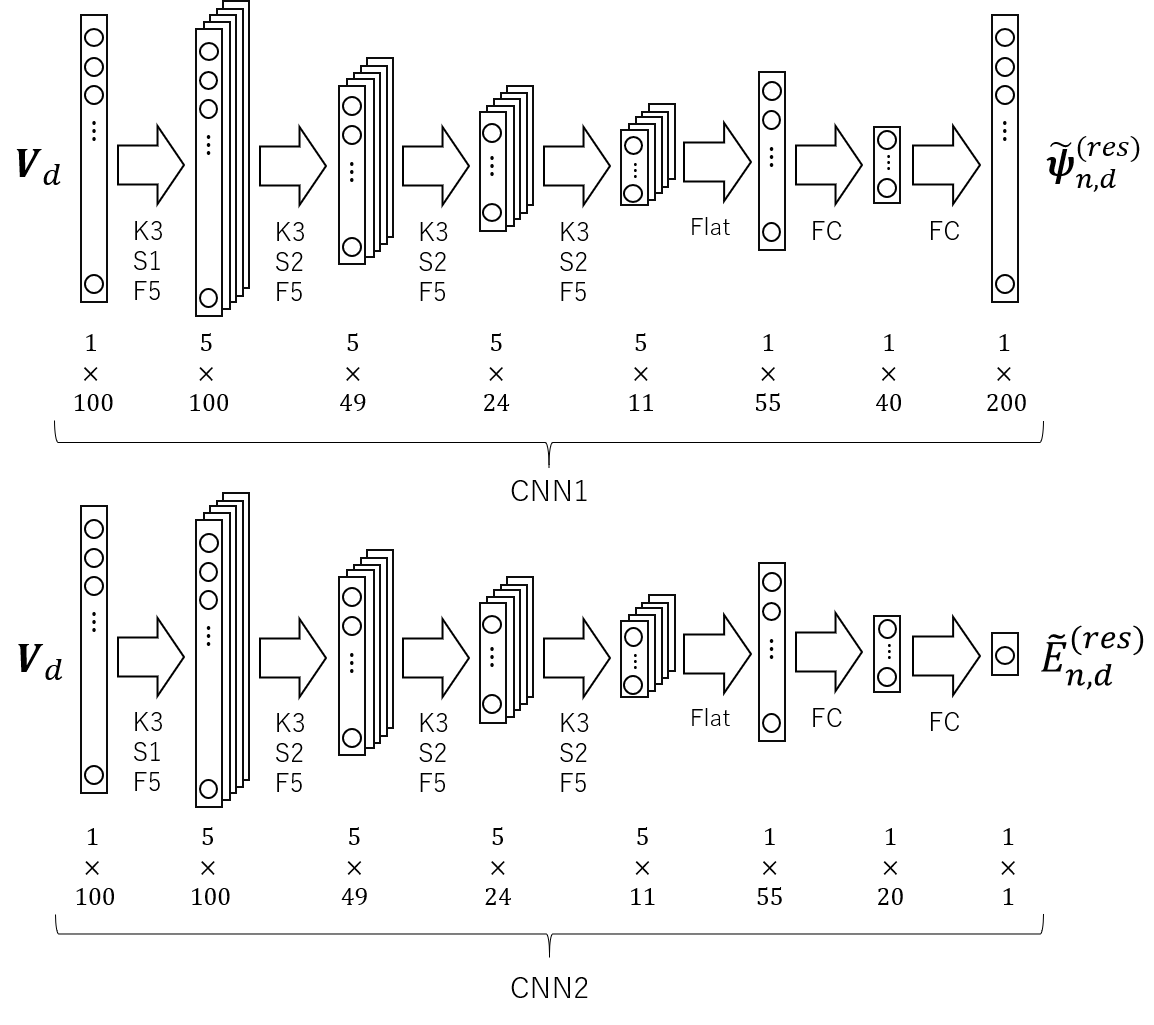}
    \caption{Two architectures of weakly-supervised learning are configured for the wave function and the energy, respectively.
    The numbers below the layers, such as $5\times100$, indicate the number of features times the number of nodes.
    The architecture of each convolutional layer is specified by an alphabet followed by an integer. 
    For example, K3, S2, and F5 mean that the window length is 3, the stride length is 2, and the number of output filters is 5.
    Feeding a perturbation potential defined on the discretized coordinate space to the CNNs, 
    we obtain output of a corresponding residual wave function and residual energy.}
    \label{fig:CNNarc}
   \end{figure}

   Two-dimensional CNN is considered to be a good architecture to deal with spatial information (e.g. image).
   Similarly, one-dimensional CNN is applied to capture the feature of spatially continuous function, such as signals \cite{1DCNN}.
   It is also reported that CNN effectively expresses the feature of one-dimensional functions using fewer parameters than a fully-connected neural network does \cite{DRN}.
   In this study, therefore, we employ two of one-dimensional CNN architectures shown in Fig. \ref{fig:CNNarc}.
   In a forward propagation process, we feed an input perturbation potential ${V}_d$ to both CNN1 and CNN2, 
   which then predict the residual wave function $\tilde{\psi}_{n,d}^{(res)}$ and the residual energy $\tilde{E}_{n,d}^{(res)}$, respectively.
   In each CNN architecture, first four hidden layers are one-dimensional convolutional layers that are supposed to extract latent features of the input.
   Each of the convolutional layers is composed of 5 filters with kernel size 3 and stride 1.
   The latent features extracted from the CNN layers are fed to fully-connected (FC) layers to obtain final outputs.
   We denote parameters in CNN1 and CNN2 as ${\bm \theta}_1$ and ${\bm \theta}_2$, respectively.
   These are optimized alternately during training, which means that one of them is fixed while optimizing the other.
   Although the architectures of CNN1 and CNN2 are independent, 
   their outputs will be considered together in the loss function that we will introduce later.
   Hence, the optimized parameters ${\bm \theta}_1$ and ${\bm \theta}_2$ are dependent on each other through the loss function.

   For numerical computation of a function in the coordinate $x\in\mathbb{R}$, we discretize $x$ into bins with an equal width,
   and treat them as the input and output nodes of CNNs.

   \subsubsection*{Loss function}
   In order for the CNNs to learn the mapping from the input perturbation potential to the corresponding residual wave function and energy,
   we employ a loss function $L$ of the following form.
   \begin{equation}
    \label{eq:loss}
     L({\bm \theta_1}, {\bm \theta}_2)
     = \frac{1}{D}\sum_d^D\left\{ \left|{H}_d\tilde{\psi}_{n,d} - \tilde{E}_{n,d}\tilde{\psi}_{n,d}\right|
     + \beta {\rm ReLU}\left[ \left|E_{n,d}^{(1)}-\tilde{E}_{n,d}^{(res)}\right|-\alpha \right] \right\},
   \end{equation}    
   where $|\cdot |$ represents Euclidean norm and $D$ is the number of  different perturbations, 
   and ReLu is the rectified linear unit.
   The predictions $\tilde{\psi}_{n,d}^{(res)}$ and $\tilde{E}_{n,d}^{(res)}$ of the CNNs depend on ${\bm \theta}_1$ and ${\bm \theta}_2$, respectively,
   and finally we obtain 
   \begin{align}
    \tilde{\psi}_{n,d}=\psi_{n}^{(0)}+\tilde{\psi}_{n,d}^{(res)},\\
    \tilde{E}_{n,d}=E_{n}^{(0)}+\tilde{E}_{n,d}^{(res)}.
   \end{align}
   It is important to note that the exact solutions for $H_{d}$ are not required in the loss function,
   and only the first perturbation energies $E_{n,d}^{(1)}$ for various perturbation potentials are used.
   This is why we call the method weakly-supervised.
   The first term of the loss function is minimized when the predicted wave functions and energies satisfy the TISE. 
   The second term is small if $\tilde{E}_{n,d}^{(res)}$ is close to the first-order perturbation energy ${E}_{n,d}^{(1)}$
   which is prepared in advance (see next section).
   The second term is added to the loss function so as to fix the predictions to the desired quantum state $n$
   and avoid them to go to another quantum state during training.
   Because of the ReLu function, the second term can be positive only when the absolute difference $\left|E_{n,d}^{(1)}-\tilde{E}_{n,d}^{(res)}\right|$ exceeds $\alpha$.
   This is essential for the weakly-supervised learning of TISE, and one of main contributions of this study.

   $\alpha$ and $\beta$ in the second term are hyperparameters. 
   $\alpha$ allows discrepancy between the residual energy and the first-order perturbation energy
   because the former needs to be close but not equal to the latter.
   $\beta$ determines the ratio between the first term and second term of the loss function.
   The performance of the method is sensitive to $\alpha$ and it has bounds (see Supplementary Information).
   Through optimization of ${\bm \theta}_1$ and ${\bm \theta}_2$ with the loss function minimized over $D$ samples in the training set,
   the CNNs are expected to produce good approximate solutions of TISE for various inputs of perturbation potentials.

   To implement the differentiation contained in the Hamiltonians, 
   we use the finite difference method (see Supplementary Information).
   \subsubsection*{Data set} 
   \label{sec:data}
   As the CNNs are trained to solve a variety of perturbation potentials,
   we need a data set composed of a variety of input perturbation potentials as well as the first-order wave functions and energies.
   A wider variety of input potentials would be preferable to both training and testing 
   because it enables the CNNs to be more flexible for unseen input perturbation potentials.
   Therefore, we consider a finite set of orthogonal functions with the size of $J$,
   \begin{equation}
    \label{eq:set}
     \mathcal{V}=\{v^{(1)},v^{(2)},\cdots,v^{(J)}\}
   \end{equation}
   and define an input perturbation potential as
   \begin{equation}
    \label{eq:pote}
     V_d = \sum_{j=1}^{J} \lambda_d^{(j)} v^{(j)},\quad \lambda_d^{(j)} \sim U(-\Lambda,\Lambda).
   \end{equation}
   Here, $\lambda_d^{(j)}$ is a random coefficient for the orthogonal function $v^{(j)}$
   and sampled from the uniform distribution with the interval $[-\Lambda,\Lambda]$.
   Hereinafter we call the set (\ref{eq:set}) ``basic perturbation set.''
   The set could be a finite number of polynomial functions, trigonometric functions, or others.
   By randomly generating $\lambda$s to obtain a sufficient number of $V_{d}$, 
   we can construct a data set that covers a wide range of the function space under the restriction of $\Lambda$ and $J$.
   The data set would be of a sufficient size for training the CNNs and testing the generalization of the trained CNNs.
   Note that the superscripts in equation (\ref{eq:pote}) are not related to the order of the perturbation,
   and $V_{d}$ and $v^{(j)}$s are the functions of $x$.

   For each input potential $V_d$, the first-order perturbation energy $E_{n,d}^{(1)}$ is necessary to calculate the loss function (\ref{eq:loss}).
   We implemented numerical integration in order to obtain these first-order perturbation corrections.
   It is noted that both the difficulty of the training and the generalization of the trained CNNs highly depend on the range of the random coefficients, $\Lambda$.
   We will show the dependence in the section of \hyperref[sec:results]{Numerical experiment and results}.
   
   \subsubsection*{Pre-training}
   As we have the first perturbation energy for $V_d$, we can implement pre-training for CNN2 by using the information. 
   To be precise, before minimizing the loss function (\ref{eq:loss}), 
   we pre-train CNN2 by minimizing $\left|E_{n,d}^{(1)} - \tilde{E}_{n,d}^{(res)}\right|$.
   In addition, if it is not costly to calculate ${\psi}_{n,d}^{(1)}$, 
   we can also pre-train CNN1 by minimizing $\left|\psi_{n,d}^{(1)} - \tilde{\psi}_{n,d}^{(res)}\right|$. 
   Using the parameters optimized by the pre-training would be a good start for the actual minimization of the loss function
   because the first-order perturbation corrections are the most dominant parts in the perturbation corrections.

  \subsection*{Errors for testing}
  After the weakly-supervised learning described above, one would like to know whether the outputs from the trained model are fairly correct or not.
  To test the validity of our method, 
  one can measure the errors in the wave function and energy averaged over randomly chosen $D$ samples of the perturbations of the form (\ref{eq:pote}) 
  with the basic perturbation set $\mathcal{V}$. 
  The errors in the wave function and energy are defined respectively by
  \begin{equation}
   \label{eq:error}
    \begin{aligned}
     \text{ERROR}^{wf}  &=\frac{1}{D}\sum_d^D \frac{\left|\psi_{n,d} - \tilde{\psi}_{n,d}\right|}{\left|\psi_{n,d}\right|} \\
     \text{ERROR}^{ene} &=\frac{1}{D}\sum_d^D \frac{\left|E_{n,d} - \tilde{E}_{n,d}\right|}{\left|E_{n,d}\right|},
    \end{aligned}
  \end{equation}
  where $\psi_{n,d}$ and $E_{n,d}$ are the exact eigen function and the eigen energy, respectively, calculated by the diagonalization of ${H}_d$.
  Note that the diagonalization of ${H}_d$ is done for testing our method and is not required in the training phase.

 \section*{Numerical experiment and results}
 \label{sec:results}
 In this section, we show experimental results of our proposed method. For the numerical experiments, 
 we consider the excited state ($n=1$) of Harmonic oscillator with perturbations.
 The results of the experiments are validated by comparing them with exact diagonalization calculations of the Hamiltonians.

  \subsection*{Harmonic oscillator and perturbations}
  To confirm the validity of the method, 
  we consider a one-dimensional harmonic oscillator (HO) as unperturbed system described by the following Hamiltonian:
  \begin{align} 
   \label{eq:HO0}
   {H}_0 = -\frac{\hslash^2}{2m}\frac{d^2}{dx^2} + \frac{m w^2 {x}^2}{2}.
  \end{align}
  We focus on a coordinate range $-1<x\leq1$ and $x_0 \equiv \sqrt{\frac{\hslash}{mw}} = 0.15$ 
  so that the wave function would be sufficiently close to zero at both ends of the range.
  For training, we choose the basic perturbation set defined in (\ref{eq:set}) in the form of trigonometric functions:
  \begin{align}
   \label{eq:Vtrain}
   \mathcal{V}^{(train)} = \{1,\sin(\pi x),\cos(\pi x),\cdots,\sin(25\pi x),\cos(25\pi x) \}
  \end{align}
  where each element has a different wave number with an integer of $1$ to $25$.
  We divide the coordinate range into $100$ bins and the maximum wave number ($25\pi$) in $\mathcal{V}^{(train)}$ is chosen 
  so that the wave length is sufficiently larger than the size of a single bin.

  \subsection*{Results of training}
\begin{figure}[tb]
    \begin{minipage}{0.48\textwidth}
     \centering
     \includegraphics[width=\textwidth]{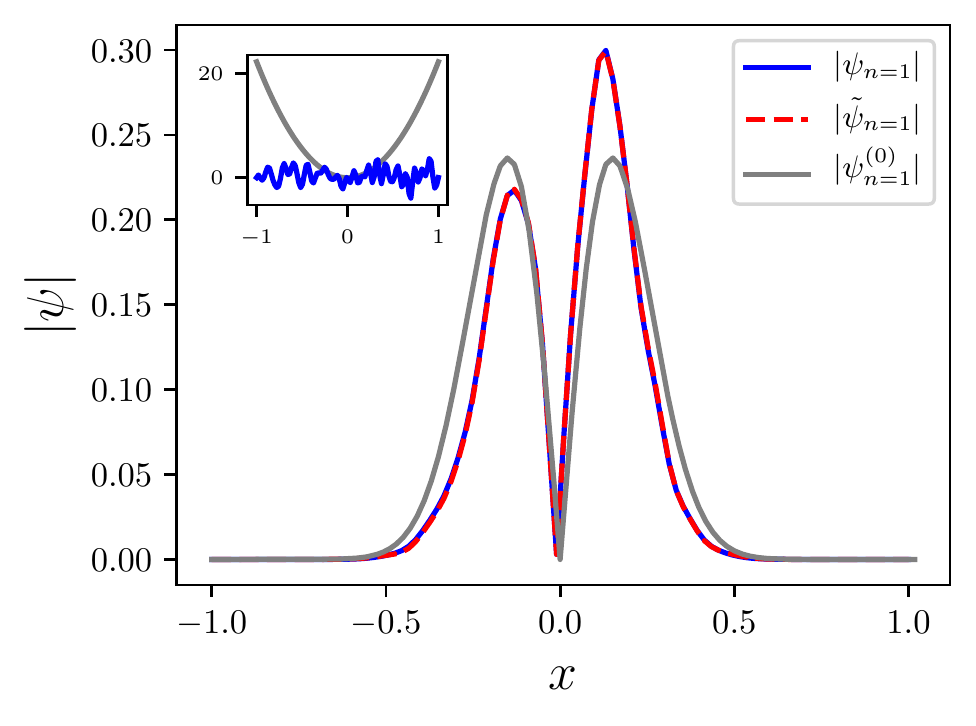}
     \caption{Absolute value of the first-excited-state wave functions of the harmonic oscillator with one of the randomly-generated 4,096 perturbation samples.
     The output of CNN1 agrees well with the wave function calculated by the exact diagonalization of the Hamiltonian.
     The unperturbed wave function is also depicted for reference. The inset shows the harmonic potential and the perturbation used for the calculation.}
     \label{fig:out_wave}
    \end{minipage}
    \hfill
    \begin{minipage}{0.48\textwidth}
     \centering
     \includegraphics[width=\textwidth]{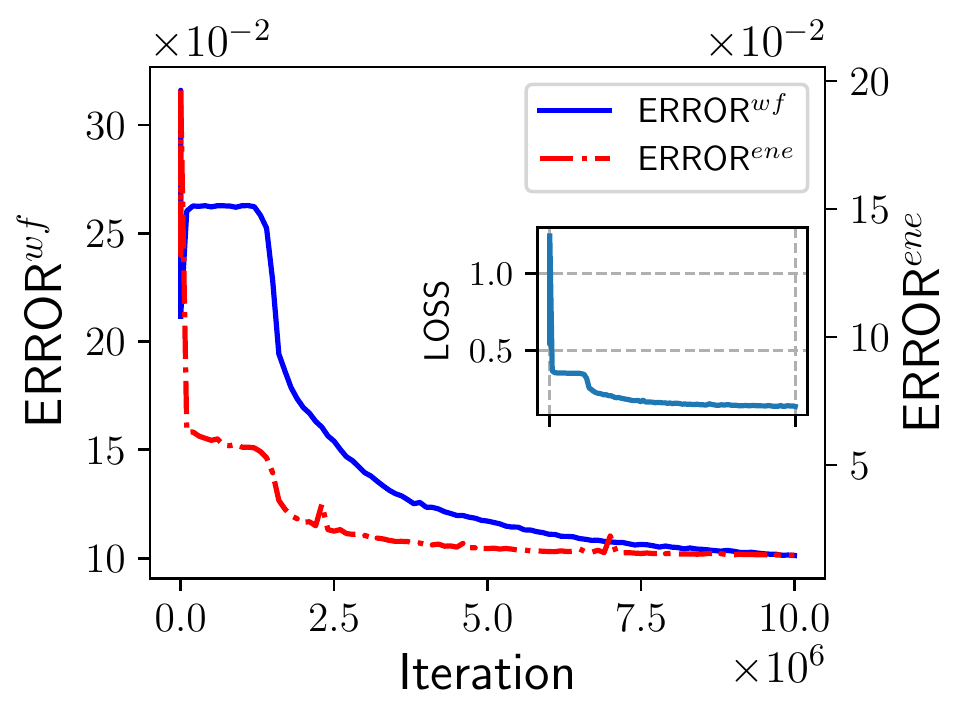}
     \caption{Errors averaged over the training set as functions of the training iteration. 
     In the training, CNN1 and CNN2 were optimized alternately. An iteration is defined as one step of the mini-batch gradient descent for both CNN1 and CNN2.
     The inset shows the loss defined in (\ref{eq:loss}). The horizontal axis of the inset is the same as the outer one.}
     \label{fig:errors_HO1}
    \end{minipage}
   \end{figure}

  Generating random coefficients $\lambda_d^{(j)} \sim U(-\Lambda_{train},\Lambda_{train})$ with $\Lambda_{train}=0.5$,
  we prepared a training set with 4,096 samples on $\mathcal{V}^{(train)}$.
  The loss function (\ref{eq:loss}) on the training set is minimized by Adam's mini-batch stochastic gradient descent \cite{Adam}.
  After the training, the optimized CNN produces an approximate wave function
  which agrees well with the exact wave function $\psi_{n,d}$ obtained by the diagonalization of ${H}_d$ (Fig. \ref{fig:out_wave}).
  The loss decreases over iterations of the optimization as shown in the inset of Fig. \ref{fig:errors_HO1}.
   We also measure the errors defined in (\ref{eq:error}) over all training samples.
   The CNNs never see the exact solutions of TISE with our weakly supervised scheme though,
   the errors shown in Fig. \ref{fig:errors_HO1} achieve the level of $0.11$ for wave functions and $0.01$ for energies, respectively.
   It should also be noted that the perturbations with $\Lambda_{train}=0.5$ can be comparable with the unperturbated potential near $x=0$ 
   as one can see from the inset of Fig. \ref{fig:out_wave}.
   It means that our method can deal with reasonably strong perturbations.
   The relationship between the value of $\Lambda_{train}$ and the errors will be discussed in the next subsection.

   Note that our method can be applied to a case where exact solution cannot be accessible
   although in this numerical experiment we prepared the exact solutions by diagonalization in order to check the validity. 
   Another advantage is that our model can be generalized for various shapes of potential once the model is well trained.
   We will discuss the generalization property with regard to input perturbation potentials in the next section.
   
   \subsection*{Results of testing}
   \label{sec:gene}
   In the above subsection, we trained the CNNs with the training set which was composed of the randomly-generated linear combination of trigonometric functions. 
   Now we test the trained model by giving it another perturbation potential. 
   One can choose any potential but here we pick up another basic perturbation set to use as test perturbation potential: 
   $\mathcal{V}^{(test)}=\{P_0,P_1,\cdots,P_{40}\}$
   consisting of Legendre polynomials (see Supplementary Information) with the random coefficients $\lambda_d^{(j)} \sim (-\Lambda_{test},\Lambda_{test})$.
   Here, we again set the maximum order of Legendre polynomials according to the size of a single bin.
   Note that the testing phase does not require any extra computation (e.g. diagonalization).
   Since our model has already been trained with the unperturbed Hamiltonian defined in (\ref{eq:HO0}), we just provide the test perturbation potentials as inputs to the model. 
   The diagonalization for the test perturbations is done only for the estimation of the errors defined in equation (\ref{eq:error}).

\begin{figure}[tb]
    \begin{minipage}{0.48\textwidth}
     \centering
     \includegraphics[width=\textwidth]{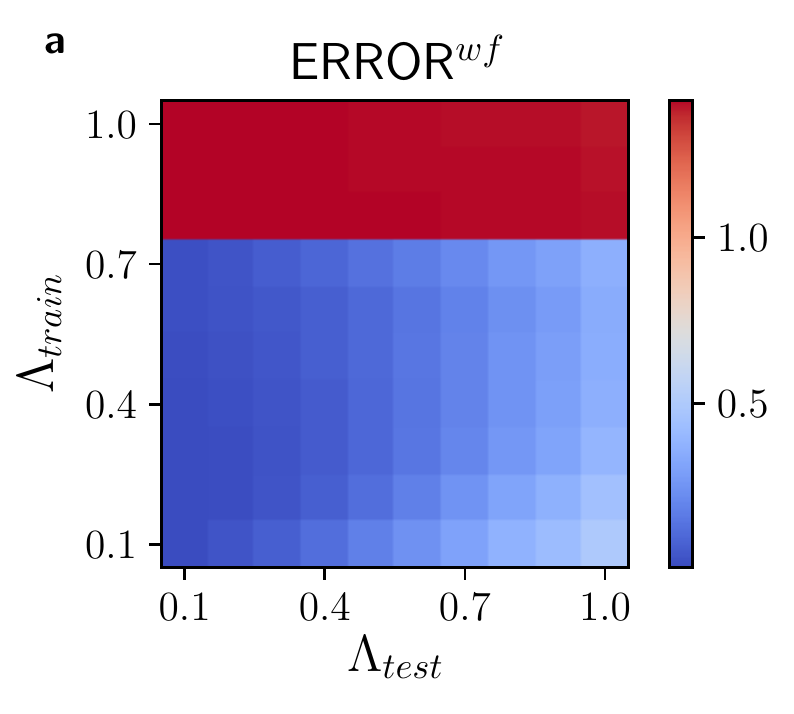}
    \end{minipage}
    \hfill
    \begin{minipage}{0.48\textwidth}
     \centering
     \includegraphics[width=\textwidth]{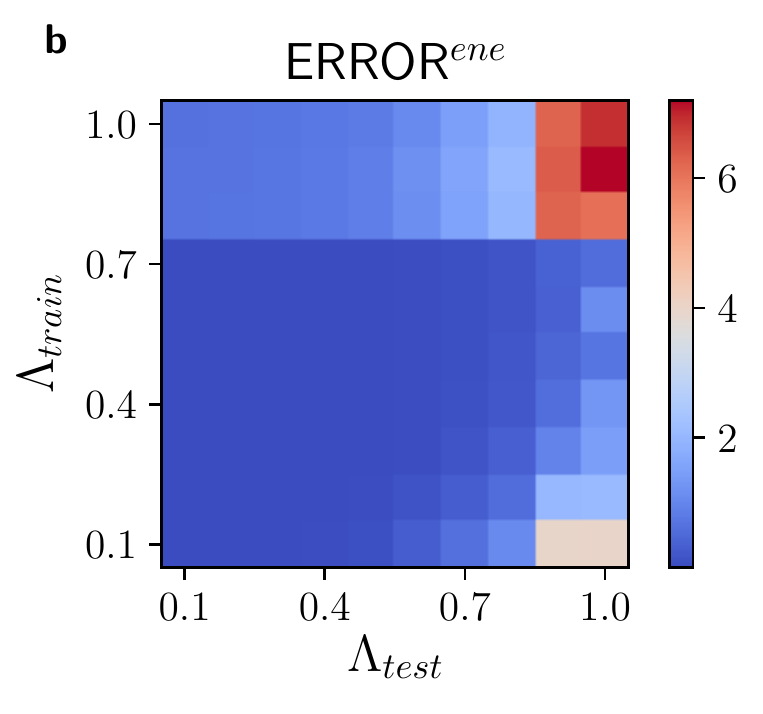}
    \end{minipage}
    \caption{The errors in \textbf{a}) the wave function and \textbf{b}) the energy defined in (\ref{eq:error}).
    After preparing the trigonometric basic perturbation sets with different values of $\Lambda_{train}$ defined in (\ref{eq:pote}),
    we constructed some CNNs each of which is trained on different $\Lambda_{train}$. 
    For trained CNNs, we measured the errors on another basic perturbation set, Legendre polynomials with $\Lambda_{test}$. }
    \label{fig:cmaps}
   \end{figure}

   As we can expect from the idea of the perturbation theory and the limitation of machine learning,
   the accuracy for a test data highly depends on the magnitude of the input perturbation potential and the quality of the training set. 
   To see those features, we train CNNs on $\mathcal{V}^{(train)}$ with different values of $\Lambda_{train}$
   and calculate the errors defined in (\ref{eq:error}) on $\mathcal{V}^{(test)}$ with different values of $\Lambda_{test}$.
   The color maps of the errors in the wave function and the energy for $\Lambda_{train}$ and $\Lambda_{test}$ 
   ranging from $0.1$ to $1.0$ with the intervals of $0.1$ are depicted in Fig. \ref{fig:cmaps}.
   First, we see obvious critical lines (around $\Lambda_{train}=0.8$) in the error maps. 
   This is caused by the energy cross between the target quantum state and a neighboring quantum state. 
   It makes training of CNNs difficult due to the limitation of the hyperparameter $\alpha$ in the loss function (see Supplementary Information).
   Another possible reason is the fixed number of the training samples of perturbation for every $\Lambda_{train}$.
   With larger $\Lambda_{train}$, we have a wider range of choices of the training samples and therefore we need more samples to maintain the accuracy of the training. 
   However since we fixed the number of the training samples, too large $\Lambda_{train}$ (beyond the critical line) caused the reduction of the accuracy.

   The figures show that the errors for small $\Lambda_{train}$ remain small, although larger than those on the training set.
   This indicates that our method can produce approximate solutions of TISE for various input potentials.
   We can see that CNNs trained on a wider range of random coefficients (i.e. larger $\Lambda_{train}$ 
   but not beyond the critical line) have a higher accuracy over $\Lambda_{test}$. 
   In other words, a wider range of the random coefficients in the training set is preferable for higher accuracy in test phase.

\section*{Discussion} 
\label{sec:dis}
We developed a new methodology of solving Schr\"{o}dinger equations using weakly-supervised deep learning.
The idea is the following. 
First, we pick up various samples of perturbations to be added to the unperturbed Hamiltonian 
and perform first-order calculation of the wave function and energy, which is usually easy. 
The obtained information of the first-order corrections is used to define the loss function. 
The machine learning model is then trained on the perturbation samples to minimize the loss function. 
The model is now ready to be used to calculate states and energies of the unperturbed Hamiltonian with an arbitrary perturbation added.
A characteristic of the method is that in the training phase it does not require exact solution (or ground truth) that can be expensive or even impossible to prepare.

As an example, we trained the model on randomly-chosen combinations of trigonometric functions as perturbations to a harmonic oscillator. After the training, 
the model was applied to the harmonic oscillator with another perturbation (linear combination of Legendre functions) 
and the resulting wave functions and energies agreed well with those obtained from exact diagonalization.

Although we conducted the numerical experiments on the first excited state of the harmonic oscillator with perturbations, 
the method can be applied to calculate any excited states of various systems if one can obtain its first order perturbation information beforehand.
In addition, our method is not limited to one-dimensional systems and easily extended to higher spatial dimensions.
It could also be an interesting future work to apply the method to time-dependent Schr\"{o}dinger equations using time-dependent perturbation theory.

\section*{Data availability}
The data that support the plots within this paper
and other findings of this study are available from the corresponding
author upon request.

\AtEndEnvironment{thebibliography}{
\bibitem{Chen2018a} 
Chen, R. T. Q., Rubanova, Y., Bettencourt, J. \& Duvenaud, D.
Neural ordinary differential equations.
Preprint at \url{http://arxiv.org/abs/1806.07366} (2018).

\bibitem{Bishop} 
C. M. Bishop.
Pattern recognition and machine learning (information science and statistics)
(New York, NY : Springer, 2006).

\bibitem{Zhu} 
X. Zhu.
Semi-supervised  learning  literature  survey.
Technical Report1530, Department of Computer Sciences, University of Wisconsin,Madison.
\url{http://digital.library.wisc.edu/1793/60444} (2005).

\bibitem{10.1093/nsr/nwx106}
Zhi-Hua, Z.
A brief introduction to weakly supervised learning.
{\it Natl. Sci. Rev.} {\bf 5}, 44–53,
DOI: \url{https://doi.org/10.1093/nsr/nwx106} (2017).

\bibitem{Adam}
Kingma, D. P. \& Ba, J. Adam: A method for stochastic optimization. 
Preprint at \url{https://arxiv.org/abs/1412.6980} (2014).
}

\bibliographystyle{naturemag-doi}
\bibliography{main}

\section*{Acknowledgements}
We would like to thank Liu Wei and Sojeong Park for proof reading of our paper and giving valuable comments.
This work was supported by a Research Fellowships of Japan Society for the Promotion of Science for Young Scientists, Grant Number 20J12472.
K. S. is also grateful to the A*STAR (Agency for Science, Technology and Research) Research Attachment Programme of Singapore for financial support.

\section*{Author contributions}
K. S. and H. K. L. designed the study and performed computer calculations. 
All authors analyzed the results and approved the final manuscript.

\section*{Competing interests}
The authors declare no competing interests.

\section*{Additional information}
Supplementary information is available in the online version of the paper. \\
Correspondence and requests for materials should be addressed to K. S.


%

\end{document}